\documentclass[12pt]{article}
\usepackage{graphicx}
\usepackage{lscape}
\usepackage{citesort}
\usepackage{amssymb}
\usepackage{appendix}
\usepackage{multirow}

\begin{document}
\def\qq{\langle \bar q q \rangle}
\def\uu{\langle \bar u u \rangle}
\def\dd{\langle \bar d d \rangle}
\def\sp{\langle \bar s s \rangle}
\def\GG{\langle g_s^2 G^2 \rangle}
\def\Tr{\mbox{Tr}}
\def\figt#1#2#3{
        \begin{figure}
        $\left. \right.$
        \vspace*{-2cm}
        \begin{center}
        \includegraphics[width=10cm]{#1}
        \end{center}
        \vspace*{-0.2cm}
        \caption{#3}
        \label{#2}
        \end{figure}
    }

\def\figb#1#2#3{
        \begin{figure}
        $\left. \right.$
        \vspace*{-1cm}
        \begin{center}
        \includegraphics[width=10cm]{#1}
        \end{center}
        \vspace*{-0.2cm}
        \caption{#3}
        \label{#2}
        \end{figure}
                }

\def\ds{\displaystyle}
\def\beq{\begin{equation}}
\def\eeq{\end{equation}}
\def\bea{\begin{eqnarray}}
\def\eea{\end{eqnarray}}
\def\beeq{\begin{eqnarray}}
\def\eeeq{\end{eqnarray}}
\def\ve{\vert}
\def\vel{\left|}
\def\ver{\right|}
\def\nnb{\nonumber}
\def\ga{\left(}
\def\dr{\right)}
\def\aga{\left\{}
\def\adr{\right\}}
\def\lla{\left<}
\def\rra{\right>}
\def\rar{\rightarrow}
\def\lrar{\leftrightarrow}
\def\nnb{\nonumber}
\def\la{\langle}
\def\ra{\rangle}
\def\ba{\begin{array}}
\def\ea{\end{array}}
\def\tr{\mbox{Tr}}
\def\ssp{{\Sigma^{*+}}}
\def\sso{{\Sigma^{*0}}}
\def\ssm{{\Sigma^{*-}}}
\def\xis0{{\Xi^{*0}}}
\def\xism{{\Xi^{*-}}}
\def\qs{\la \bar s s \ra}
\def\qu{\la \bar u u \ra}
\def\qd{\la \bar d d \ra}
\def\qq{\la \bar q q \ra}
\def\gGgG{\la g^2 G^2 \ra}
\def\q{\gamma_5 \not\!q}
\def\x{\gamma_5 \not\!x}
\def\g5{\gamma_5}
\def\sb{S_Q^{cf}}
\def\sd{S_d^{be}}
\def\su{S_u^{ad}}
\def\sbp{{S}_Q^{'cf}}
\def\sdp{{S}_d^{'be}}
\def\sup{{S}_u^{'ad}}
\def\ssp{{S}_s^{'??}}

\def\sig{\sigma_{\mu \nu} \gamma_5 p^\mu q^\nu}
\def\fo{f_0(\frac{s_0}{M^2})}
\def\ffi{f_1(\frac{s_0}{M^2})}
\def\fii{f_2(\frac{s_0}{M^2})}
\def\O{{\cal O}}
\def\sl{{\Sigma^0 \Lambda}}
\def\es{\!\!\! &=& \!\!\!}
\def\ap{\!\!\! &\approx& \!\!\!}
\def\ar{&+& \!\!\!}
\def\ek{&-& \!\!\!}
\def\kek{\!\!\!&-& \!\!\!}
\def\cp{&\times& \!\!\!}
\def\se{\!\!\! &\simeq& \!\!\!}
\def\eqv{&\equiv& \!\!\!}
\def\kpm{&\pm& \!\!\!}
\def\kmp{&\mp& \!\!\!}
\def\mcdot{\!\cdot\!}
\def\erar{&\rightarrow&}


\def\simlt{\stackrel{<}{{}_\sim}}
\def\simgt{\stackrel{>}{{}_\sim}}


\renewcommand{\textfraction}{0.2}    
\renewcommand{\topfraction}{0.8}

\renewcommand{\bottomfraction}{0.4}
\renewcommand{\floatpagefraction}{0.8}
\newcommand\mysection{\setcounter{equation}{0}\section}

\def\baeq{\begin{appeq}}     \def\eaeq{\end{appeq}}
\def\baeeq{\begin{appeeq}}   \def\eaeeq{\end{appeeq}}
\newenvironment{appeq}{\beq}{\eeq}
\newenvironment{appeeq}{\beeq}{\eeeq}
\def\bAPP#1#2{
 \markright{APPENDIX #1}
 \addcontentsline{toc}{section}{Appendix #1: #2}
 \medskip
 \medskip
 \begin{center}      {\bf\LARGE Appendix #1 :}{\quad\Large\bf #2}
\end{center}
 \renewcommand{\thesection}{#1.\arabic{section}}
\setcounter{equation}{0}
        \renewcommand{\thehran}{#1.\arabic{hran}}
\renewenvironment{appeq}
  {  \renewcommand{\theequation}{#1.\arabic{equation}}
     \beq
  }{\eeq}
\renewenvironment{appeeq}
  {  \renewcommand{\theequation}{#1.\arabic{equation}}
     \beeq
  }{\eeeq}
\nopagebreak \noindent}

\def\eAPP{\renewcommand{\thehran}{\thesection.\arabic{hran}}}

\renewcommand{\theequation}{\arabic{equation}}
\newcounter{hran}
\renewcommand{\thehran}{\thesection.\arabic{hran}}

\def\bmini{\setcounter{hran}{\value{equation}}
\refstepcounter{hran}\setcounter{equation}{0}
\renewcommand{\theequation}{\thehran\alph{equation}}\begin{eqnarray}}
\def\bminiG#1{\setcounter{hran}{\value{equation}}
\refstepcounter{hran}\setcounter{equation}{-1}
\renewcommand{\theequation}{\thehran\alph{equation}}
\refstepcounter{equation}\label{#1}\begin{eqnarray}}


\newskip\humongous \humongous=0pt plus 1000pt minus 1000pt
\def\caja{\mathsurround=0pt}


\title{
         {\Large
                 {\bf Strong Coupling Constants of Decuplet Baryons with Vector Mesons 
                 }
         }
      }

\author{\vspace{1cm}\\
{\small T. M. Aliev$^a$ \thanks {e-mail:
taliev@metu.edu.tr}~\footnote{permanent address: Institute of
Physics, Baku, Azerbaijan}\,\,, K. Azizi$^b$ \thanks {e-mail:
kazizi@dogus.edu.tr}\,\,, M. Savc{\i}$^a$ \thanks
{e-mail: savci@metu.edu.tr}} \\
{\small $^a$ Physics Department, Middle East Technical University,
06531 Ankara, Turkey} \\
{\small $^b$ Physics Division, Faculty of Arts and
Sciences, Do\u gu\c s University,  Ac{\i}badem-} \\
{\small Kad{\i}k\"oy, 34722 Istanbul, Turkey} }

\date{}

\begin{titlepage}
\maketitle
\thispagestyle{empty}

\begin{abstract}
We provide a comprehensive study of  strong coupling constants of decuplet baryons with light nonet  vector mesons in the framework of  light
cone QCD sum rules. Using the symmetry arguments, we argue that all coupling constants entering the calculations can be expressed in terms of only one invariant function  even
if the $SU(3)_f$ symmetry breaking effects are taken into account. We estimate the order of  $SU(3)_f$ symmetry violations, which are automatically considered by the employed approach.
\end{abstract}

~~~PACS number(s): 11.55.Hx, 13.30.-a, 13.30.Eg, 14.20.Jn 
\end{titlepage}

\section{Introduction}

Theoretically, the baryon-baryon-meson coupling constants are fundamental objects as they can provide useful  information on the low energy QCD, baryon-baryon interactions and scattering of
mesons from baryons.    In other words, their values calculated in QCD can    render important constraints
in constructing baryon-baryon as well as baryon-meson potentials. 
 They can help us to better analyze the results of existing experiments on the meson-nucleon, nucleon-hyperon and
hyperon-hyperon interactions held in different  centers, such as MAMI, MIT, Bates, BNL and Jefferson Laboratories.

Calculation of  the baryon-meson  coupling constants using  the fundamental theory of QCD is highly desirable. However, such interactions occur in a region  very far from the perturbative regime and the 
fundamental QCD Lagrangian
is not suitable for calculation of  these coupling constants. Therefore,  we need some non-perturbative approaches. QCD sum rules \cite{R10501} is one of the most powerful and applicable tools in this respect. It is based on the QCD Lagrangian,
 hence the problem of deriving the baryon-meson coupling from QCD sum rules is clearly of importance, both
as a fundamental test of QCD and of the applied  non-perturbative approach.

 In the present work, we calculate the strong coupling constants of decuplet baryons with light nonet  vector mesons   in
 the framework of the light cone QCD sum rules  \cite{R10502}. Applying  the symmetry arguments, we derive all related coupling constants in terms of
 only one universal function even if 
 $SU(3)_f$ symmetry breaking effects are encountered. One of the main advantage of the approach used during this work is that it automatically includes the $SU(3)_f$ symmetry breaking effects.  Calculation of these coupling constants is also very important
for understanding the dynamics of light vector mesons and their
electroproduction  off the decuplet baryons. Note that the strong coupling constants of the octet and decuplet baryons with pseudoscalar mesons as well as octet baryons with vector mesons   have been studied within the same framework in \cite{R10503,R10504,R10505,R10506,R10506kazem}.

The  layout of the paper is  as follows. In section 2, using the symmetry relations, sum rules for the strong
coupling constants of the light nonet  vector mesons with decuplet
baryons are obtained in the framework of light cone QCD sum rules (LCSR).   In section 3, we numerically analyze the coupling constants of the light nonet  vector mesons with decuplet
baryons,  estimate the order of  $SU(3)_f$ symmetry violations and discuss the obtained results.

\section{Sum rules for the strong coupling constants of the the light nonet  vector mesons with decuplet baryons}
In this part, we derive LCSR for the coupling constants of the  light nonet  vector mesons with decuplet
baryons and show how it is possible to express all couplings entering the calculations in terms of only one universal function. In $SU(3)_f$ symmetry the  interaction Lagrangian can be written as:
  \bea
\label{e10501} {\cal L}_{\rm int} = g \varepsilon_{ijk}
(\bar{\cal D}_{\ell m}^j)^{\mu} ({\cal D}^{m\ell k})_\mu  \partial^n{
V}_n^i + \mbox{\rm h.c.}~, \eea
where  ${\cal D}$ and
$ V$ stand for   decuplet baryons and vector
mesons, respectively. To obtain the sum rules for coupling constants, we start considering the following correlation function, which is the main  building block in QCD sum rules:

 \bea \label{e10502} \Pi_{\mu\nu} = i \int d^4x
e^{ipx} \lla { V}(q) \vel {\cal T}\left\{ \eta_\mu(x)
\bar{\eta}_\nu(0) \right\} \ver 0 \rra~, \eea where ${ V}(q)$ corresponds to
the light mesons with momentum $q$, $\eta_\mu$ is
the interpolating currents for decuplet  baryons and ${\cal T}$ is the time ordering operator. To obtain sum rules for the coupling constants, we will calculate the correlation function in following two different 
 ways:
\begin{itemize}
 \item in phenomenological side, the correlation function is obtained in terms of hadronic parameters saturating in by a tower of hadrons with the same quantum numbers as the interpolating currents.
\item in theoretical or QCD side, the correlation function is calculated by means of operator product expansion (OPE) in deep Euclidean
region, where $-p^2 \rar \infty$ and $-(p+q)^2 \rar \infty$, in terms of quark and gluon degrees of
freedom. By the help of the OPE the short and large distance effects are separated. The short range effects are calculated using the perturbation theory, whereas the long distance contributions are parametrized 
 in terms of DA's of the light nonet  vector mesons.
\end{itemize}
Finally, to get the sum rules, we  equate these two representations of the correlation functions through dispersion relation and apply Borel transformation with respect to the variables $(p+q)^2$ and $p^2$ to suppress the contribution of the higher states
 and continuum.
Before starting calculations of the correlation function in physical or theoretical sides let us introduce the interpolating currents of the decuplet baryons.  The interpolating currents  creating the
 decuplet baryons   can be written in a  compact form as:
\bea
\label{e10503}\eta_\mu \es A \varepsilon^{abc} \left\{
(q_1^{aT} C \gamma_\mu q_2^b) q_3^c + (q_2^{aT} C \gamma_\mu q_3^b)
q_1^c + (q_3^{aT} C \gamma_5 q_1^b) q_2^c\right\}~, \eea where
$a,b$ and $c$ are the color indices and
$C$ is the charge conjugation operator. The values of normalization
constant $A$  and the $q_1$, $q_2$ and $q_3$ quarks  are represented in Table 1.

\begin{table}[h]

\renewcommand{\arraystretch}{1.3}
\addtolength{\arraycolsep}{-0.5pt}
\small
$$
\begin{array}{|l|c|c|c|c|}
\hline \hline
 & A & q_1 & q_2 & q_3 \\  \hline
 \Sigma^{\ast 0} & \sqrt{2/3}  & u & d & s  \\
 \Sigma^{\ast +} & \sqrt{1/3}  & u & u & s  \\
 \Sigma^{\ast -} & \sqrt{1/3}  & d & d & s  \\
 \Delta^{++}     &       1/3   & u & u & u  \\
 \Delta^{+}      & \sqrt{1/3}  & u & u & d  \\
 \Delta^{0}      & \sqrt{1/3}  & d & d & u  \\
 \Delta^{-}      &       1/3   & d & d & d  \\
 \Xi^{\ast 0}    & \sqrt{1/3}  & s & s & u  \\
 \Xi^{\ast -}    & \sqrt{1/3}  & s & s & d  \\
 \Omega^{-}      &       1/3   & s & s & s  \\
\hline \hline
\end{array}
$$
\caption{The values of $A$ and the quark flavors $q_1$, $q_2$
and $q_3$ for decuplet baryons.}
\renewcommand{\arraystretch}{1}
\addtolength{\arraycolsep}{-1.0pt}

\end{table}

  As we already noted, the phenomenological side of correlation function is obtained  inserting a full set of hadrons with quantum numbers of $\eta_\mu$ and 
isolating the ground state  baryons as:
 \bea \label{e10506}
\Pi_{\mu\nu}(p,q) = { \lla 0 \vel \eta_\mu \ver {\cal D}(p_2) \rra \lla {\cal
D}(p_2) {V}(q) \ve {\cal D}(p_1) \rra \lla {\cal D}(p_1) \vel
\bar \eta_\nu \ver 0 \rra \over (p_2^2-m_{{\cal D}2}^2) (p_1^2 - m_{{\cal
D}1}^2) } + \cdots~, \eea where $m_{{\cal
D}1}$ and $m_{{\cal
D}2}$ are masses of the initial and final state decuplet baryons with momentum
 $p_1=p+q$ and $p_2=p$, respectively and $\cdots$
represents  contribution of the  higher states and  continuum. 

To proceed, we need to know the the matrix element of the interpolating current between vacuum and decuplet state as well as the transition matrix element. The $\lla {\cal D}(p_1) \vel \eta_\mu \ver 0 \rra$ is defined 
in terms of the residue $\lambda_{\cal D}$ as:
\bea
\label{e10507}
\lla 0 \vel \eta_\mu \ver {\cal D}(p) \rra \es \lambda_{\cal D}
u_\mu(p)~, \eea where $u_\mu$ is the Rarita--Schwinger
spinor. 
The transition matrix element, $\lla {\cal D}(p_2) V(q) \ve {\cal D}(p_1)
\rra$,  is  parameterized in terms of coupling form factors $g_1$, $g_2$,  $g_3$ and $g_4$ as:
\bea
\label{e10508} \lla  {\cal D}(p_2) V(q) \ve {\cal D}(p_1)
\rra &=& \bar{u}_{\alpha}(p_2)\left\{g^{\alpha\beta} \left[\rlap/{\varepsilon}g_1+2p.\varepsilon \frac{g_2}{(m_{{\cal
D}1}+m_{{\cal
D}2})}\right]\right.\nnb \\ &+& \left.\frac{q^{\alpha}q^{\beta}}{(m_{{\cal
D}1}+m_{{\cal
D}2})^2}\left[\rlap/{\varepsilon}g_3+2p.\varepsilon \frac{g_4}{(m_{{\cal
D}1}+m_{{\cal
D}2})}\right]\right\} u_\beta (p_1) ~, \eea
 Using Eqs. (\ref{e10507}) and (\ref{e10508})
into (\ref{e10506}) and performing summation over spins of the  decuplet baryons using, \bea \label{e10509}
\sum_s u_\mu(p,s) \bar{u}_\nu(p,s) \es (\rlap/{p} + m_{\cal
D}) \Bigg\{ g_{\mu\nu} - {\gamma_\mu \gamma_\nu \over3} - {2
p_{\mu} p_{\nu} \over 3 m_{\cal D}^2} + {p_{\mu}\gamma_\nu -
p_{\nu} \gamma_\mu \over 3 m_{\cal D}} \Bigg\}~, \eea in principle, one can
obtain the final  expression for the phenomenological side of the
correlation function. However, there is two problems which we should overcome: all existing structures are not independent and the interpolating current for decuplet
baryons couples also to unwanted
spin--1/2 states, i.e., 
 \bea \label{e10510} \lla 0 \vel \eta_\mu
\ver 1/2(p) \rra = (A \gamma_\mu + B p_{\mu}) u(p)~, \eea
exists and has nonzero value.
 Multiplying both sides
of Eq. (\ref{e10510}) with $\gamma_\mu$ and using $\eta_\mu
\gamma^\mu = 0$, we get $B=-4A/m_{1/2}$. From this relation, we see that to remove the contribution of the unwanted spin--1/2 states,  we should 
eliminate the terms proportional to  $\gamma_\mu$ at the left   $\gamma_\nu$ at the right and also terms containing  $p_{2\mu}$ and $p_{1\nu}$. For this aim and also to get independent structures, we order the 
Dirac matrices as $\gamma_\mu\rlap/{\varepsilon}\rlap/{q}\rlap/{p}\gamma_\nu$ and set the terms containing the contribution of spin--1/2 particles to zero.
After this  procedure, we obtain the final expression for the phenomenological side as: 
\bea \label{e10511} \Pi_{\mu\nu} &=&
{ \lambda_{{\cal D}1} \lambda_{{\cal D}2} \over
[m_{{\cal D}1}^2-(p+q)^2] [m_{{\cal D}2}^2-p^2]} \left\{2(\varepsilon.p)g_{\mu\nu}\rlap/{q}\left[g_1+g_2\frac{m_{{\cal
D}2}}{(m_{{\cal
D}1}+m_{{\cal
D}2})}\right]\right. \nnb \\ 
&-&2(\varepsilon.p)g_{\mu\nu}\rlap/{q}\rlap/{p}\frac{g_2}{(m_{{\cal
D}1}+m_{{\cal
D}2})}+q_\mu q_\nu\rlap/{\varepsilon}\rlap/{q}\rlap/{p}\frac{g_3}{(m_{{\cal
D}1}+m_{{\cal
D}2})^2} \nnb \\ 
&-&2\left.(\varepsilon.p)q_\mu q_\nu\rlap/{q}\rlap/{p}\frac{g_4}{(m_{{\cal
D}1}+m_{{\cal
D}2})^3}+
\mbox{\rm other structures}\right\}~, \eea  where, to obtain sum rules for coupling constants, we will choose the structures, $(\varepsilon.p)g_{\mu\nu}\rlap/{q}$, 
$(\varepsilon.p)g_{\mu\nu}\rlap/{q}\rlap/{p}$, $q_\mu q_\nu\rlap/{\varepsilon}\rlap/{q}\rlap/{p}$ and $(\varepsilon.p)q_\mu q_\nu\rlap/{q}\rlap/{p}$ for form factors
 $g_1+g_2$,  $g_2$, $g_3$ and $g_4$, respectively.

In this part, before calculation of the QCD side of the aforementioned correlation function, we would like to present the relations
between invariant functions for the coefficients of the selected structures and show how we can express all coupling constants in terms of only one universal function. The main advantage of this
approach used below is that it takes into account $SU(3)_f$
symmetry violating effects, automatically. Following the works \cite{R10503,R10504,R10505,R10506,R10506kazem}, we start
 considering the transition, $\Sigma^{\ast 0} \rar
\Sigma^{\ast 0} \rho^0$ whose invariant function correspond to each coupling $g_1$, $g_2$, $g_3$ and $g_4$  can
formally be written as: \bea \label{e10512}
\Pi^{\Sigma^{\ast 0} \rar \Sigma^{\ast 0}\rho^0} = g_{\rho^0 \bar{u}u}
\Pi_1(u,d,s) +
 g_{\rho^0 \bar{d}d} \Pi_1^\prime(u,d,s) + g_{\rho^0 \bar{s}s} \Pi_2(u,d,s)~,
\eea
where, from the interpolating current of the $\rho^0$ meson we have
$g_{\rho^0 \bar{u}u} = - g_{\rho^0 \bar{d}d} = 1/\sqrt{2}$, and
$g_{\rho^0 \bar{s}s} = 0$. In the above relation, the invariant functions $\Pi_1$, $\Pi_1^\prime$ and $\Pi_2$
refer to the radiation of $\rho^0$ meson from $u$, $d$ and $s$ quarks, respectively, and we formally define them as:
\bea
\label{e10514}
\Pi_1(u,d,s) \es \lla \bar{u}u \vel \Sigma^{\ast 0}
\Sigma^{\ast 0} \ver 0 \rra~, \nnb \\
\Pi_1^\prime(u,d,s) \es \lla \bar{d}d \vel \Sigma^{\ast 0}
\Sigma^{\ast 0} \ver 0 \rra~, \nnb \\
\Pi_2(u,d,s) \es \lla \bar{s}s \vel \Sigma^{\ast 0}
\Sigma^{\ast 0} \ver 0 \rra~.
\eea
The interpolating currents of the $\Sigma^{\ast 0}$ is  symmetric under  $u \lrar d$, hence
$\Pi_1^\prime(u,d,s) = \Pi_1(d,u,s)$ and  Eq.
(\ref{e10512}) immediately yields:
\bea
\label{e10515}
\Pi^{\Sigma^{\ast 0} \rar \Sigma^{\ast 0}\rho^0} = {1\over \sqrt{2}}
[\Pi_1(u,d,s) - \Pi_1(d,u,s)]~,
\eea
where, in the $SU(2)_f$ symmetry limit it vanishes. Now, we proceed considering the invariant function describing the transition, $\Sigma^{\ast +} \rar
\Sigma^{\ast +}\rho^0$. It can be obtained from Eq. (\ref{e10512}) by
replacing $d \rar u$ and using the fact that
$\Sigma^{\ast 0} (d \rar u) = \sqrt{2} \Sigma^{\ast +}$. As a result we get,
 \bea \label{e10516} 4 \Pi_1(u,u,s) =  2 \lla \bar{u}u \vel
\Sigma^{\ast +} \Sigma^{\ast +} \ver 0 \rra~, \eea where the coefficient 4 in the left side comes from the fact that the  $\Sigma^{\ast +}$
contains two $u$ quarks and there are 4 possibilities  for $\rho^0$ meson
to be radiated from the $u$ quark. Using Eq. (\ref{e10512}) and
considering the fact that $\Sigma^{\ast +}$ does not contain
$d$ quark, we obtain \bea \label{e10517} \Pi^{\Sigma^{\ast +} \rar
\Sigma^{\ast +}\rho^0} =  \sqrt{2} \Pi_1(u,u,s)~. \eea From the similar way, the invariant
function describing the $\Sigma^{\ast -} \rar \Sigma^{\ast -}\rho^0$
is obtained from $\Sigma^{\ast 0} \rar \Sigma^{\ast -}\rho^0$ replacing $u \rar d$ in Eq. (\ref{e10512})
and taking into account  $\Sigma^{\ast 0} (u \rar d) = \sqrt{2} \Sigma^{\ast -}$, i.e.,
\bea \label{e10518} \Pi^{\Sigma^{\ast -} \rar
\Sigma^{\ast -}\rho^0} = - \sqrt{2} \Pi_1(d,d,s)~. \eea 

Our next task is to expand the approach to include the  $\Delta$
baryons. The invariant function for the $\Delta^+ \rar \Delta^+ \rho^0$
transition can be obtained from the $\Sigma^{\ast+} \rar \Sigma^{\ast+} \rho^0$
transition. From the interpolating currents it is clear that, $\eta^{\Delta^+}_\mu = \eta^{\Sigma^{\ast +}}_\mu (s
\rar d)$. Using this fact, we obtain:
\bea
\label{e10519}
\Pi^{\Delta^+ \rar \Delta^+ \rho^0} \es  \Big[ g_{\rho^0\bar{u}u} \lla \bar{u}u \vel
\Sigma^{\ast +} \Sigma^{\ast +} \ver 0 \rra (s \rar d) +
g_{\rho^0\bar{s}s} \lla \bar{s}s \vel
\Sigma^{\ast +} \Sigma^{\ast +} \ver 0 \rra (s \rar d) \Big]\nnb \\
\es \sqrt{2} \Pi_1(u,u,d) - {1\over \sqrt{2}} \Pi_2 (u,u,d)~,
\eea
but our calculations show that,
\bea \label{kaz1} \Pi_2 (u,u,d)=\Pi_1 (d,u,u)~, \eea 
hence,\bea
\label{e10519prime1}
\Pi^{\Delta^+ \rar \Delta^+ \rho^0} 
\es \sqrt{2} \Pi_1(u,u,d) - {1\over \sqrt{2}} \Pi_1 (d,u,u)~.
\eea
Similar to the above relations, our calculations lead also to the following relations for the couplings of remaining decuplet baryons with $\rho^0$ meson:
\bea
\label{e10519prime2}
\Pi^{\Delta^{++} \rar \Delta^{++} \rho^0} 
\es  {3\over \sqrt{2}} \Pi_1 (u,u,u)~,
\eea
\bea
\label{e10519prime3}
\Pi^{\Delta^{-} \rar \Delta^{-} \rho^0} 
\es - {3\over \sqrt{2}} \Pi_1 (d,d,d)~,
\eea
\bea
\label{e10520}
\Pi^{\Delta^0 \rar \Delta^0 \rho^0} \es- \sqrt{2} \Pi_1(d,d,u) +
{1\over \sqrt{2}} \Pi_1 (u,d,d)~, \eea
\bea
\Pi^{\Xi^{\ast 0} \rar \Xi^{\ast 0} \rho^0} \es {1\over \sqrt{2}}
\Pi_1(u,s,s)~, \eea
\bea
\Pi^{\Xi^{\ast -} \rar \Xi^{\ast -} \rho^0} \es {-1\over \sqrt{2}}
\Pi_1(d,s,s)~.
\eea

Up to here,  we considered the  neutral $\rho$ meson case. Now, we go on considering the  relations among the invariant functions correspond to the  charged $\rho$ meson, for instance $\Sigma^{\ast 0} 
\rar \Sigma^{\ast+} \rho^-$. For this aim, we start considering 
the matrix element $\lla \bar{d}d \vel \Sigma^{\ast 0}
\Sigma^{\ast 0} \ver 0 \rra$, where $d$ quark from each $\Sigma^{\ast 0}$  constitutes the final $\bar{d}d$ state, and the remaining
$u$ and $s$  are  spectator quarks. From the similar way, in the
matrix element $\lla \bar{u}d \vel \Sigma^{\ast+} \Sigma^{\ast 0} \ver 0 \rra$,
$d$ quark from $\Sigma^{\ast 0}$ and $u$ quark from $\Sigma^{\ast+}$ form
$\bar{u}d$ state and the remaining $u$ and $s$ quarks remain also as spectators.
As a result one expects that these two matrix elements  should be proportional. Our calculations support this expectation and lead to the following relation:
\bea
\label{e10121}
\Pi^{\Sigma^{\ast 0} \rar \Sigma^{\ast+} \rho^-} \es \lla \bar{u}d \vel \Sigma^{\ast+}
\Sigma^{\ast 0} \ver 0 \rra = \sqrt{2} \lla \bar{d}d \vel \Sigma^{\ast0}
\Sigma^{\ast 0} \ver 0 \rra \nnb \\
\es  \sqrt{2} \Pi_1(d,u,s)~.
\eea
$\Sigma^{\ast 0} 
\rar \Sigma^{\ast-} \rho^+$ invariant function is obtained exchanging the  $u \lrar d$ in the above relation, i.e.,
\bea
\label{e10122}
\Pi^{\Sigma^{\ast 0} \rar \Sigma^{\ast-} \rho^+} \es \lla \bar{d}u \vel \Sigma^{\ast-}  
\Sigma^{\ast 0} \ver 0 \rra = \sqrt{2} \lla \bar{u}u \vel \Sigma^{\ast0}
\Sigma^{\ast 0} \ver 0 \rra \nnb \\
\es \sqrt{2} \Pi_1(u,d,s)~.
\eea
We obtain the
following relations among other invariant functions involving charged $\rho$
meson using the similar arguments and calculations:
\bea
\label{e10123}
\Pi^{\Sigma^{\ast -} \rar \Sigma^0 \rho^-} \es 
\sqrt{2} \Pi_1(u,d,s)~, \eea
\bea
\Pi^{\Xi^{\ast -} \rar \Xi^0 \rho^-} \es 
 \Pi_1(d,s,s)=\Pi_1(u,s,s)~, \eea
\bea
\Pi^{\Delta^+ \rar \Delta^{++}  \rho^-} \es \sqrt{3}  \Pi_1(u,u,u)~, \eea
\bea
\Pi^{\Delta^0 \rar \Delta^+  \rho^-} \es 2 \Pi_1(u,u,d)~, \eea
\bea
\Pi^{\Delta^- \rar \Delta^0  \rho^-} \es \sqrt{3} \Pi_1(d,d,d)~, \eea
\bea
\Pi^{\Sigma^{\ast +} \rar \Sigma^{\ast0} \rho^+} \es 
\sqrt{2} \Pi_1(d,u,s)~, \eea
\bea     
\Pi^{\Xi^{\ast 0} \rar \Xi^{\ast-} \rho^+} \es
 \Pi_1(d,s,s)~, \eea
\bea
\Pi^{\Delta^+ \rar \Delta^0 \rho^+} \es  2 \Pi_1(d,d,u)~,
\eea
\bea
\Pi^{\Delta^{++} \rar \Delta^+ \rho^+} \es  \sqrt{3} \Pi_1(d,u,u)~,
\eea
\bea
\Pi^{\Delta^{0} \rar \Delta^- \rho^+} \es  \sqrt{3} \Pi_1(u,d,d)~.
\eea
The remaining relations among the invariant functions involving other light nonet vector mesons,  $K^{\ast0,\pm}$, $\bar{K}^{\ast0}$, $\omega$  and $\phi$
are represented in Appendix A.
The above relations as well as those presented in the Appendix A show how we can express all strong coupling constants of the decuplet baryons to light vector mesons in terms of one universal function,
 $\Pi_1$. 

Now, we focus our attention to calculate this invariant function in terms of the QCD degrees of freedom. As it is seen from the interpolating currents of the decuplet baryons previously shown, one can
 describe all transitions in terms of 
 $\Sigma^{\ast 0} \rar \Sigma^{\ast 0} \rho^0$, so we will calculate the invariant function $\Pi_1$ only for this transition. From QCD or theoretical side, the correlation function can be calculated 
in deep Euclidean region, where $-p^2 \rar \infty$, $-(p+q)^2 \rar \infty$, via operator product expansion (OPE) in terms of e distribution amplitudes (DA's) of the light vector mesons and light quarks
 propagators. Therefore, to proceed, we need to know the 
 expression of the light quark propagator as well as the matrix elements of the nonlocal 
operators $\bar{q}(x_1) \Gamma q^\prime (x_2)$ and $\bar{q}(x_1) G_{\mu\nu} 
q^\prime (x_2)$ between the vacuum and the vector meson states. Here, 
 $\Gamma$ refers the Dirac matrices correspond to the case under
consideration and $G_{\mu\nu}$ is the gluon field strength tensor. Up to twist--4 accuracy, the 
matrix elements $\lla V(q) \vel \bar{q}(x) \Gamma q(0) \ver 0 \rra$ and 
$\lla V(q) \vel \bar{q}(x) G_{\mu\nu} q(0) \ver 0 \rra$ are determined in
terms of the distribution amplitudes (DA's) of the vector mesons 
\cite{R10110,R10111,R10112}. For simplicity, we present these nonlocal matrix elements  in
Appendix B. The expressions for DA's of the light vector mesons are also given in \cite{R10110,R10111,R10112}.

The light quark propagator used in our calculations is:
\bea
\label{e10125}
S_q(x) \es {i \rlap/x \over 2 \pi^2 x^4} - {m_q \over 4 \pi^2 x^2} -
{\lla\bar{q}q\rra \over 12} \Bigg(1 - {i m_q \over 4} \rlap/x \Bigg) -
{x^2 \over 192} m_0^2 \lla\bar{q}q\rra  \Bigg(1 - {i m_q \over 6} 
\rlap/x \Bigg) \nnb \\
\ek i g_s \int_0^1 du \Bigg\{ {\rlap/x \over 16 \pi^2 x^2} G_{\mu\nu} (ux)
\sigma^{\mu\nu} - u x^\mu G_{\mu\nu} (ux) \gamma^\nu {i \over 4 \pi^2 x^2}
\nnb \\
\ek {i m_q \over 32 \pi^2} G_{\mu\nu} (ux) \sigma^{\mu\nu} \Bigg[ \ln \Bigg(
{- x^2 \Lambda^2 \over 4} \Bigg) + 2 \gamma_E \Bigg] \Bigg\}~,
\eea 
where $\gamma_E$ is the Euler gamma and  $\Lambda$ is a scale parameter which it is chosen
 as a factorization scale, i.e., $\Lambda = (0.5-1.0)~GeV$  \cite{R10113}. Using the expression of the light quark propagator and 
the DA's of the light vector mesons, the theoretical or QCD  side of the correlation 
function is obtained. Equating the coefficients of the structures, $(\varepsilon.p)g_{\mu\nu}\rlap/{q}$, 
$(\varepsilon.p)g_{\mu\nu}\rlap/{q}\rlap/{p}$, $q_\mu q_\nu\rlap/{\varepsilon}\rlap/{q}\rlap/{p}$ and $(\varepsilon.p)q_\mu q_\nu\rlap/{q}\rlap/{p}$ from both representations of the correlation
function in phenomenological and theoretical sides  and
applying Borel transformation with respect to the variables $p^2$ and $(p+q)^2$  to suppress the
contributions of the higher states and continuum, we get the sum rules for strong coupling constants of the  vector
mesons to  decuplet baryons,
\bea
\label{e10126}
g_1 +{g_2m_{{\cal D}2} \over(m_{{\cal D}1}+m_{{\cal D}2})}\es  {1\over 2\lambda_{{\cal D}1} \lambda_{{\cal D}2}}e^{{m^2_{{\cal D}1} \over M_1^2} + 
{m^2_{{\cal D}2} \over M_2^2} + {m_V^2 \over{M_1^2 + M_2^2}}} \, \Pi_1^{(1)}~,
\nnb \\
g_2 \es -{(m_{{\cal D}1}+m_{{\cal D}2})\over 2\lambda_{{\cal D}1} \lambda_{{\cal D}2}}e^{{m^2_{{\cal D}1} \over M_1^2} + 
{m^2_{{\cal D}2} \over M_2^2} + {m_V^2 \over{M_1^2 + M_2^2}}} \, \Pi_1^{(2)}~, \nnb \\
g_3 \es {(m_{{\cal D}1}+m_{{\cal D}2})^2\over \lambda_{{\cal D}1} \lambda_{{\cal D}2}}e^{{m^2_{{\cal D}1} \over M_1^2} + 
{m^2_{{\cal D}2} \over M_2^2} + {m_V^2 \over{M_1^2 + M_2^2}}} \,\Pi_1^{(3)}~,\nnb \\
g_4 \es-{(m_{{\cal D}1}+m_{{\cal D}2})^3\over 2\lambda_{{\cal D}1} \lambda_{{\cal D}2}}e^{{m^2_{{\cal D}1} \over M_1^2} + 
{m^2_{{\cal D}2} \over M_2^2} + {m_V^2 \over{M_1^2 + M_2^2}}} \,\Pi_1^{(4)}~,
\eea
where, $M_1^2$ and $M_2^2$ are Borel parameters corresponding to the initial and final baryon channels, respectively and the functions, $\Pi_1^{(i)}$ which are functions of the QCD degrees of freedom, 
continuum threshold 
 as well as
mass,  decay constant  and
DA's of the  light vector mesons have very lengthy 
expressions and for this reason, we do not present their explicit expressions here. It should be noted here that, the masses of the initial and final baryons
are close to each other, so we will set
$M_1^2=M_2^2=2 M^2$. From the sum rules  for the strong couplings of the vector mesons to decuplet baryons in Eq. (\ref{e10126}), it is clear that we also 
need the 
residues of decuplet baryons. These residues  are obtained using the  two--point correlation functions  in
\cite{R10508,R10518,R10519} (see also \cite{R10506kazem}). 

\section{Numerical analysis}

In this section, we numerically analyze  the sum rules of the strong coupling constants of  the light nonet
vector mesons  with decuplet baryons and discuss our results. The sum rules for the couplings, $g_1$, $g_2$, $g_3$ and  $g_4$ depict that the main input 
parameters  are the vector meson 
DA's. The DA's of the vector mesons which  are calculated in    
\cite{R10110,R10111,R10112} include the leptonic constants, $f_V$ and 
$f_V^T$,  the twist--2 and twist--3 parameters, $a_i^\parallel$,
$a_i^\perp$,$\zeta_{3V}^\parallel$, $\tilde{\lambda}_{3V}^\parallel$,
$\tilde{\omega}_{3V}^\parallel$, $\kappa_{3V}^\parallel$,
$\omega_{3V}^\parallel$, $\lambda_{3V}^\parallel$, $\kappa_{3V}^\perp$,
$\omega_{3V}^\perp$, $\lambda_{3V}^\perp$, and twist--4 parameters 
$\zeta_4^\parallel$, $\tilde{\omega}_4^\parallel$, $\zeta_4^\perp$,
 $\tilde{\zeta}_4^\perp$, $\kappa_{4V}^\parallel$, $\kappa_{4V}^\perp$. The values of all these parameters are
given in Tables 1 and 2 in \cite{R10112}.
 The values of the remaining
parameters entering  the sum rules are:  $\langle 0|\frac{1}{\pi}\alpha_{s}G^{2}|0\rangle=(0.012
\pm 0.004)~ GeV^{4}$ \cite{Ioffeb}, $\lla \bar{u} u\rra = \lla \bar{d} d \rra = -
(0.24\pm 0.01)^3~GeV^3$, $\lla \bar{s} s\rra =0.8 \lla \bar{u} u\rra $ \cite{Ioffeb}, $m_0^2=(0.8 \pm 0.2)~GeV^2$ \cite{R10508}, $ m_{s}(2~GeV)=(111 \pm 6)~MeV$ at 
 $\Lambda_{QCD}=330~MeV$ \cite{Dominguez}. In numerical calculations, we set $m_u=m_d=0$.

The sum rules for the  coupling constants contain also two auxiliary parameters, Borel mass parameter $M^2$ and
continuum threshold $s_0$.  Therefore, we should find  working regions of these parameters, 
where the results of  coupling 
constants are reliable. In the reliable regions, the coupling constants are weakly depend on the auxiliary parameters.
The upper limit of the Borel parameter, $M^2$ is found  demanding
that the contribution of the higher states and continuum  should be less than, say 40\% of the total value of the same correlation
function. The lower limit of $M^2$ is found  requiring that the
contribution of the highest term with the power of $1/M^2$ be 20--25\% less than
that of the highest power of $M^2$.  As a result, we obtain the working region, $1~GeV^2\leq M^2\leq1.5~GeV^2$ for the Borel mass parameter. The continuum threshold is also not completely arbitrary but depends 
 on the energy of the first excited state with the same quantum numbers. Our calculations lead to the working  region,
$3~GeV^2\leq s_0\leq5~GeV^2$ for   the continuum threshold.


\begin{table}[h]

\renewcommand{\arraystretch}{1.3}
\addtolength{\arraycolsep}{-0.5pt}
\small
$$
\begin{array}{r@{\rar}lr@{\pm}lr@{\pm}l}
\hline \hline
\multicolumn{2}{c}{\mbox{\rm Channel}} &  \multicolumn{2}{c}{\mbox{\rm~~~~~~~ $g_1$} } &\multicolumn{2}{c}{\mbox{\rm ~~~~~~~~~$g_1$(SU(3)) }} \\ \hline
\Delta^+        & \Delta^+ \rho^0  & $~~~~~~~~~~$ -4.4&0.9  & $~~~~~~~~~~$-4.4&0.9\\
\Omega^-        & \Xi^{\ast -} K^{\ast0}      & -23.5&4.6  &  -13.2&2.5\\
\Sigma^{\ast 0} & \Sigma^{\ast 0} \phi  &  -8.0&1.7 & -7.3&1.5\\
\Sigma^{\ast 0} & \Xi^{\ast 0} K^{\ast0}     & -18.5&3.8  &  -10.8&2.2\\
\Sigma^{\ast -} & \Sigma^{\ast -} \rho^0  &  9.1&2.0 & 8.8&1.8\\
 \Xi^{\ast 0}    & \Xi^{\ast 0} \rho^0    &   -4.8&1.2  &  -4.4&0.9\\
\Xi^{\ast -}    & \Sigma^{\ast -} K^{\ast0}    &   -26.0&5.4  &  -15.2&3.2\\
\hline \hline
\end{array}
$$
\caption{Coupling constant $g_1$ of light vector mesons with decuplet baryons.}
\renewcommand{\arraystretch}{1}
\addtolength{\arraycolsep}{-1.0pt}
\label{results}
\end{table}

\begin{table}[h]

\renewcommand{\arraystretch}{1.3}
\addtolength{\arraycolsep}{-0.5pt}
\small
$$
\begin{array}{r@{\rar}lr@{\pm}lr@{\pm}l}
\hline \hline
\multicolumn{2}{c}{\mbox{\rm Channel}} &  \multicolumn{2}{c}{\mbox{\rm~~~~~~~ $g_2$} } &\multicolumn{2}{c}{\mbox{\rm ~~~~~~~~~$g_2$(SU(3)) }} \\ \hline
\Delta^+        & \Delta^+ \rho^0  & $~~~~~~~~~~$ 2.45&0.50  & $~~~~~~~~~~$2.45&0.50\\
\Omega^-        & \Xi^{\ast -} K^{\ast0}      & 7.7&1.6  &  7.2&1.4\\
\Sigma^{\ast 0} & \Sigma^{\ast 0} \phi  &  2.5&0.5 & 3.6&0.8\\
\Sigma^{\ast 0} & \Xi^{\ast 0} K^{\ast0}     & 5.4&1.1  &  5.9&1.2\\
\Sigma^{\ast -} & \Sigma^{\ast -} \rho^0  &  -5.55&1.20 & -4.85&0.95\\
 \Xi^{\ast 0}    & \Xi^{\ast 0} \rho^0    &   3.21&0.64  &  2.44&0.48\\
\Xi^{\ast -}    & \Sigma^{\ast -} K^{\ast0}    &   7.7&1.5  &  8.4&1.8\\
\hline \hline
\end{array}
$$
\caption{Coupling constant $g_2$ of light vector mesons with decuplet baryons.}
\renewcommand{\arraystretch}{1}
\addtolength{\arraycolsep}{-1.0pt}
\label{results}
\end{table}
\newpage
\begin{table}[h]

\renewcommand{\arraystretch}{1.3}
\addtolength{\arraycolsep}{-0.5pt}
\small
$$
\begin{array}{r@{\rar}lr@{\pm}lr@{\pm}l}
\hline \hline
\multicolumn{2}{c}{\mbox{\rm Channel}} &  \multicolumn{2}{c}{\mbox{\rm~~~~~~~ $g_3$} } &\multicolumn{2}{c}{\mbox{\rm ~~~~~~~~~$g_3$(SU(3)) }} \\ \hline
\Delta^+        & \Delta^+ \rho^0  & $~~~~~~~~~~$ 10.4&2.4  & $~~~~~~~~~~$10.4&2.4\\
\Omega^-        & \Xi^{\ast -} K^{\ast0}      & 39.0&8.0  &  26.0&5.4\\
\Sigma^{\ast 0} & \Sigma^{\ast 0} \phi  &  17.5&3.6 & 14.0&3.2\\
\Sigma^{\ast 0} & \Xi^{\ast 0} K^{\ast0}     & 27.4&5.6  &  21.0&4.0\\
\Sigma^{\ast -} & \Sigma^{\ast -} \rho^0  &  -24.0&4.6 & -21.0&4.2\\
 \Xi^{\ast 0}    & \Xi^{\ast 0} \rho^0    &   14.0&2.8  &  10.5&2.3\\
\Xi^{\ast -}    & \Sigma^{\ast -} K^{\ast0}    &   38.5&7.6  &  29.6&6.2\\
\hline \hline
\end{array}
$$
\caption{Coupling constant $g_3$ of light vector mesons with decuplet baryons.}
\renewcommand{\arraystretch}{1}
\addtolength{\arraycolsep}{-1.0pt}
\label{results}
\end{table}
\begin{table}[h]

\renewcommand{\arraystretch}{1.3}
\addtolength{\arraycolsep}{-0.5pt}
\small
$$
\begin{array}{r@{\rar}lr@{\pm}lr@{\pm}l}
\hline \hline
\multicolumn{2}{c}{\mbox{\rm Channel}} &  \multicolumn{2}{c}{\mbox{\rm~~~~~~~ $g_4$} } &\multicolumn{2}{c}{\mbox{\rm ~~~~~~~~~$g_4$(SU(3)) }} \\ \hline
\Delta^+        & \Delta^+ \rho^0  & $~~~~~~~~~~$ -4.2&1.6  & $~~~~~~~~~~$-4.2&1.6 \\
\Omega^-        & \Xi^{\ast -} K^{\ast0}      & -19.5&6.5  &  -9.0&3.0\\
\Sigma^{\ast 0} & \Sigma^{\ast 0} \phi  &  -8.5&2.8 & -5.5&1.8\\
\Sigma^{\ast 0} & \Xi^{\ast 0} K^{\ast0}     & -12.4&4.2  &  -7.5&2.4\\
\Sigma^{\ast -} & \Sigma^{\ast -} \rho^0  &  10.5&3.6 & 8.4&2.7\\
 \Xi^{\ast 0}    & \Xi^{\ast 0} \rho^0    &   -7.0&2.4  &  -4.0&1.5\\
\Xi^{\ast -}    & \Sigma^{\ast -} K^{\ast0}    &   -17.5&5.6  &  -10.2&3.2\\
\hline \hline
\end{array}
$$
\caption{Coupling constant $g_4$ of light vector mesons with decuplet baryons.}
\renewcommand{\arraystretch}{1}
\addtolength{\arraycolsep}{-1.0pt}
\label{results}
\end{table}
As an example, the dependence of the couplings $g_1$, $g_2$, $g_3$ and $g_4$ only for couplings of $\rho^0$ meson to $\Delta^{ +}$  baryon are shown in  Figs. (1)--(4) at 
different values of the continuum threshold. From these figures, we observe  that the couplings show good stability in the ``working" region 
of $M^2$. Obviously, the coupling constants  also weakly depend on the continuum threshold  $s_0$. The results of the strong couplings  $g_1$, $g_2$, $g_3$ and $g_4$ extracted from these figures and the similar analysis 
 for the strong coupling of the other members of the light nonet vector mesons with decuplet baryons are presented in Tables 1, 2, 3, and 4, respectively. Beside the general results, these Tables also include 
 the predictions of the 
 ̄$SU(3)_f$ symmetry on the strong coupling constants. The result of the ̄$SU(3)_f$ symmetry are obtained setting the $m_s=m_u=m_d=0$, $\lla \bar{s} s\rra = \lla \bar{u} u\rra=\lla \bar{d}d\rra $, $m_V=m_\rho$ and
 $m_{\cal D}=m_\Delta$.
Note that,  in these Tables, we show  only those
 couplings  which could not be obtained by
the $SU(2)$ symmetry rotations.
The errors presented in these Tables include  the uncertainties coming from
the variation of auxiliary parameters, $M^2$ and  $s_0$  as well as uncertainties coming from the input parameters.

A quick running  into the Tables 1--4 are resulted in:

\begin{itemize}
\item For all strong couplings, $g_1$, $g_2$, $g_3$ and $g_4$, the channels having large number of strange quarks  show overall a large $SU(3)_f$ symmetry violation comparing to those with small number of s-quark. This is reasonable and is in agreement with our expectations. 
\item The maximum $SU(3)_f$ symmetry violation for $g_1$ is 44\% and belongs to the $\Omega^-\rightarrow \Xi^{\ast -} K^{\ast0} $ channel. The maximum violation of this symmetry  for  $g_3$ and $g_4$ which also belong 
to the same channel are 33\% and 53\%, respectively. However, the channel $\Sigma^{\ast 0} \rightarrow \Sigma^{\ast 0} \phi$ shows the maximum $SU(3)_f$ symmetry violation for $g_2$ with 30\%.
\item The uncertainties on the values of the $g_1$, $g_2$ and  $g_3$ are small comparing with that of $g_4$. This is because of the fact that the  $g_1$, $g_2$ and  $g_3$ show a good stability with respect to the 
auxiliary parameters in  working regions in comparison with  $g_4$.

\end{itemize}

In conclusion, we studied the   strong coupling constants of the decuplet baryons with light nonet  vector mesons in the framework of  light
cone QCD sum rules. We expressed all coupling constants entering the calculations in terms of only one universal function  even
if the $SU(3)_f$ symmetry breaking effects are taken into account. We estimated the order of  $SU(3)_f$ symmetry violations. The main advantage of the approach used in the present work is that it 
takes  into account the $SU(3)_f$ symmetry breaking effects automatically and we don't need to define another invariant function. The obtained results on the  strong coupling constants of decuplet baryons with
 light nonet  vector mesons can help us to understand  the dynamics of light vector mesons and their
electroproduction  off the decuplet baryons.

\newpage

\bAPP{A}{}

In this appendix, we present the relations among the correlation functions
involving $K^\ast$, $\omega$ and  $\phi$   mesons.

\begin{itemize}
\item Vertices involving $K^\ast$ meson.
\end{itemize}
 
\baeeq
\label{e101ap02}
\Pi^{\Delta^+ \rar \Sigma^{\ast0} K^{\ast +}} \es \sqrt{2} 
 \Pi_1(s,u,d) ~, \nnb \\
\Pi^{\Delta^0 \rar \Sigma^{\ast-} K^{\ast +}} \es  \Pi_1(s,d,d)~, \nnb \\
\Pi^{\Sigma^{\ast +} \rar \Xi^{\ast0} K^{\ast +}} \es 2 \Pi_1(s,s,u )~, \nnb \\
\Pi^{\Sigma^{\ast 0} \rar \Xi^{\ast-} K^{\ast +}} \es \sqrt{2} \Pi_1(u,d,s )~, \nnb \\
\Pi^{\Delta^{++} \rar \Sigma^{\ast +} K^{\ast +}} \es  \sqrt{3} \Pi_1(u,u,u)~, \nnb \\
\Pi^{\Xi^{\ast0} \rar \Omega^{\ast -} K^{\ast +}} \es  \sqrt{3} \Pi_1(s,s,s)~, \nnb \\
\Pi^{\Sigma^{\ast 0} \rar \Delta^+ K^{\ast -}} \es \sqrt{2} \Pi_1(s,u,d ), \nnb \\
\Pi^{\Omega^- \rar \Xi^{\ast0} K^{\ast -}} \es  \sqrt{3} \Pi_1(s,s,s )~, \nnb \\
\Pi^{\Sigma^{\ast -} \rar \Delta^0 K^{\ast -}} \es   \Pi_1(s,d,d)~, \nnb \\
\Pi^{\Xi^{\ast 0} \rar \Sigma^{\ast+} K^{\ast -}} \es  2 \Pi_1(u,u,s)~, \nnb \\
\Pi^{\Xi^{\ast -} \rar \Sigma^{\ast0} K^{\ast -}} \es \sqrt{2} \Pi_1(u,d,s )~, \nnb \\
\Pi^{\Sigma^{\ast +} \rar \Delta^{++} K^{\ast -}} \es \sqrt{3} \Pi_1(u,u,u )~, \nnb \\
\Pi^{\Xi^{\ast 0} \rar \Sigma^{\ast0} K^{\ast 0}} \es \sqrt{2} \Pi_1(d,u,s )~, \nnb \\
\Pi^{\Xi^{\ast -} \rar \Sigma^{\ast-} K^{\ast 0}} \es 2 \Pi_1(s,s,d )~, \nnb \\
\Pi^{\Sigma^{\ast 0} \rar \Delta^0 K^{\ast 0}} \es \sqrt{2} \Pi_1(s,d,u )~, \nnb \\
\Pi^{\Omega^- \rar \Xi^{\ast-} K^{\ast 0}} \es  \sqrt{3} \Pi_1(s,s,s)~, \nnb \\
\Pi^{\Sigma^{\ast +} \rar \Delta^+ K^{\ast 0}} \es  \Pi_1(s,u,u)~, \nnb \\
\Pi^{\Sigma^{\ast -} \rar \Delta^- K^{\ast 0}} \es  \sqrt{3}\Pi_1(s,d,d)~, \nnb \\
\Pi^{\Sigma^{\ast 0} \rar \Xi^{\ast 0} \bar{K}^{\ast 0}} \es \sqrt{2} \Pi_1(d,u,s )~, \nnb \\
\Pi^{\Delta^- \rar \Sigma^{\ast-} \bar{K}^{\ast 0}} \es  \sqrt{3} \Pi_1(s,d,d)~, \nnb \\
\Pi^{\Sigma^{\ast -} \rar \Xi^{\ast-} \bar{K}^{\ast 0}} \es  2 \Pi_1(s,s,d )~, \nnb \\
\Pi^{\Delta^0 \rar \Sigma^{\ast0} \bar{K}^{\ast 0}} \es  \sqrt{2}  
\Pi_1(s,d,u) ~, \nnb \\
\Pi^{\Delta^+ \rar \Sigma^{\ast+} \bar{K}^{\ast 0}} \es  \Pi_1(s,u,u)~,\nnb \\
\Pi^{\Omega^- \rar \Xi^{\ast-} \bar{K}^{\ast 0}} \es  \sqrt{3} \Pi_1(s,s,s)~.
\eaeeq

\begin{itemize}
\item Vertices involving $\omega$ meson.
\end{itemize}

\baeeq
\label{e101ap03}
\Pi^{\Sigma^{\ast 0} \rar \Sigma^{\ast 0} \omega} \es {1 \over \sqrt{2}}
[\Pi_1(u,d,s) + \Pi_1(d,u,s) ]~, \nnb \\
\Pi^{\Sigma^{\ast +} \rar \Sigma^{\ast+} \omega} \es  \sqrt{2} \Pi_1(u,u,s )~, \nnb \\
\Pi^{\Sigma^{\ast -} \rar \Sigma^{\ast-} \omega} \es \sqrt{2} \Pi_1(d,d,s )~, \nnb \\
\Pi^{\Delta^+ \rar \Delta^+ \omega} \es { 1 \over \sqrt{2}}
\Pi_1(d,u,u) + \sqrt{2}\Pi_1(u,u,d) ~, \nnb \\
\Pi^{\Delta^{++} \rar \Delta^{++} \omega} \es{3\sqrt{2} \over 2} \Pi_1(u,u,u) ~, \nnb \\
\Pi^{\Delta^{-} \rar \Delta^{-} \omega} \es {3\sqrt{2} \over 2 }\Pi_1(d,d,d) ~, \nnb \\
\Pi^{\Delta^0 \rar \Delta^0 \omega} \es \sqrt{2}
 \Pi_1(d,d,u) +{1 \over 2 }\Pi_1(u,d,d) ]~, \nnb \\
\Pi^{\Xi^{\ast 0} \rar \Xi^{\ast 0} \omega} \es{ 1 \over \sqrt{2}} \Pi_1(u,s,s)~, \nnb \\
\Pi^{\Xi^{\ast -} \rar \Xi^{\ast -} \omega} \es {1 \over \sqrt{2}} \Pi_1(d,s,s)~.
\eaeeq
\begin{itemize}
\item Vertices involving 
$\phi$ meson.
\end{itemize}
\baeeq
\label{e101ap033333}
\Pi^{\Sigma^{\ast 0} \rar \Sigma^{\ast 0} \phi} \es 
 [\Pi_1(s,d,u) ~, \nnb \\
\Pi^{\Sigma^{\ast +} \rar \Sigma^{\ast +} \phi} \es   \Pi_1(s,u,u)~, \nnb \\
\Pi^{\Sigma^{\ast -} \rar \Sigma^{\ast -} \phi} \es \Pi_1(s,d,d)~, \nnb \\
\Pi^{\Xi^{\ast 0} \rar \Xi^{\ast 0} \phi} \es 2 \Pi_1(s,s,u )~, \nnb \\
\Pi^{\Xi^{\ast -} \rar \Xi^{\ast -} \phi} \es  2 \Pi_1(s,s,d )~.
\eaeeq

\eAPP

\newpage
\bAPP{B}{}

In this appendix we present the DA's of the vector mesons
appearing in the matrix elements $\lla V(q) \vel \bar{q}(x) \Gamma q(0) \ver
0 \rra$ and $\lla V(q) \vel \bar{q}(x) G_{\mu\nu} q(0) \ver 0 \rra$,
up to twist--4 accuracy \cite{R10110,R10111,R10112}:

\baeeq
\label{e101apb01}
\lla V(q,\lambda) \vel \bar{q}_1(x) \gamma_\mu q_2(0) \ver 0 \rra \es
f_V m_V \Bigg\{ {\varepsilon^\lambda \mcdot x \over q\mcdot x} q_\mu 
\int_0^1 du e^{i \bar{u} q\mcdot x} \Bigg[ \phi_\parallel (u) + {m_V^2 x^2 \over 16}
A_\parallel (u) \Bigg] \nnb \\
\ar \Bigg( \varepsilon_\mu^\lambda - q_\mu {\varepsilon^\lambda \mcdot x \over q\mcdot x} \Bigg) \int_0^1
du e^{i \bar{u} q\mcdot x} g_\perp^v (u) \nnb \\
\ek {1\over 2} x_\mu { \varepsilon^\lambda \mcdot x
\over (q\mcdot x)^2} m_V^2 \int_0^1 du e^{i \bar{u} q\mcdot x} \Big[ g_3 (u) +
\phi_\parallel (u) - 2 g_\perp^v (u) \Big] \Bigg\}~, \nnb \\ \nnb \\
\lla V(q,\lambda) \vel \bar{q}_1(x) \gamma_\mu \gamma_5 q_2(0) \ver 0 \rra
\es - {1 \over 4} \epsilon_\mu^{\nu\alpha\beta} \varepsilon^\lambda q_\alpha
x_\beta f_V m_V \int_0^1 du e^{i \bar{u} q\mcdot x} g_\perp^a (u)~, 
\nnb \\ \nnb \\
\lla V(q,\lambda) \vel \bar{q}_1(x) \sigma_{\mu\nu} q_2(0) \ver 0 \rra \es  
- i f_V^T \Bigg\{ (\varepsilon_\mu^\lambda q_\nu - \varepsilon_\nu^\lambda
q_\mu ) \int_0^1 du e^{i \bar{u} q\mcdot x}\Bigg[\phi_\perp (u) + {m_V^2
x^2 \over 16} A_\perp (u) \Bigg] \nnb \\
\ar { \varepsilon^\lambda \mcdot x \over (q\mcdot x)^2} (q_\mu x_\nu - q_\nu
x_\mu) \int_0^1 du e^{i \bar{u} q\mcdot x} \Bigg[h_\parallel^t - {1\over 2}
\phi_\perp - {1\over 2} h_3 (u) \Bigg] \nnb \\
\ar {1\over 2} (\varepsilon_\mu^\lambda x_\nu - \varepsilon_\nu^\lambda
x_\mu) {m_V^2 \over q \mcdot x} \int_0^1 du e^{i \bar{u} q\mcdot x}
\Big[h_3(u) - \phi_\perp (u) \Big] \Bigg\}~, \nnb \\ \nnb \\
\lla V(q,\lambda) \vel \bar{q}_1(x) \sigma_{\alpha\beta} g G_{\mu\nu}(u x) q_2(0) \ver 0 \rra \es  
f_V^T m_V^2 { \varepsilon^\lambda \mcdot x \over 2 q\mcdot x} \Big[q_\alpha q_\mu
g_{\beta\nu}^\perp - q_\beta q_\mu g_{\alpha\nu}^\perp - q_\alpha q_\nu g_{\beta\mu}^\perp
+ q_\beta q_\nu g_{\alpha\mu}^\perp \Big] \nnb \\
\cp \int {\cal D}\alpha_i e^{i(\alpha_{\bar{q}}
+ u \alpha_g) q\mcdot x} {\cal T}(\alpha_i) \nnb \\ 
\ar f_V^T m_V^2 \Big[q_\alpha \varepsilon_\mu^\lambda g_{\beta\nu}^\perp - q_\beta
\varepsilon_\mu^\lambda g_{\alpha\nu}^\perp - q_\alpha
\varepsilon_\nu^\lambda g_{\beta\mu}^\perp
+ q_\beta \varepsilon_\nu^\lambda g_{\alpha\mu}^\perp \Big] \nnb \\
\cp \int {\cal D}\alpha_i e^{i(\alpha_{\bar{q}}
+ u \alpha_g) q\mcdot x} {\cal T}_1^{(4)}(\alpha_i) \nnb \\
\ar f_V^T m_V^2 \Big[q_\mu \varepsilon_\alpha^\lambda g_{\beta\nu}^\perp -
q_\mu \varepsilon_\beta^\lambda g_{\alpha\nu}^\perp - q_\nu
\varepsilon_\alpha^\lambda g_{\beta\mu}^\perp
+ q_\nu \varepsilon_\beta^\lambda g_{\alpha\mu}^\perp \Big] \nnb \\
\cp \int {\cal D}\alpha_i e^{i(\alpha_{\bar{q}}
+ u \alpha_g) q\mcdot x} {\cal T}_2^{(4)}(\alpha_i) \nnb \\
\ar {f_V^T m_V^2 \over q \mcdot x} \Big[q_\alpha q_\mu \varepsilon_\beta^\lambda
x_\nu - q_\beta q_\mu \varepsilon_\alpha^\lambda x_\nu -
q_\alpha q_\nu \varepsilon_\beta^\lambda x_\mu +
q_\beta q_\nu \varepsilon_\alpha^\lambda x_\mu \nnb \\
\cp \int {\cal D}\alpha_i e^{i(\alpha_{\bar{q}}
+ u \alpha_g) q\mcdot x} {\cal T}_3^{(4)}(\alpha_i) \nnb \\
\ar {f_V^T m_V^2 \over q \mcdot x} \Big[q_\alpha q_\mu \varepsilon_\nu^\lambda
x_\beta - q_\beta q_\mu \varepsilon_\nu^\lambda x_\alpha -
q_\alpha q_\nu \varepsilon_\mu^\lambda x_\beta +
q_\beta q_\nu \varepsilon_\mu^\lambda x_\alpha \nnb \\
\cp \int {\cal D}\alpha_i e^{i(\alpha_{\bar{q}}
+ u \alpha_g) q\mcdot x} {\cal T}_4^{(4)}(\alpha_i)~, \nnb \\ \nnb \\
\lla V(q,\lambda) \vel \bar{q}_1(x) g_s G_{\mu\nu} (ux) q_2(0) \ver 0 \rra \es
-i f_V^T m_V (\varepsilon_\mu^\lambda q_\nu - \varepsilon_\nu^\lambda q_\mu)
\int {\cal D}\alpha_i e^{i(\alpha_{\bar{q}}
+ u \alpha_g) q\mcdot x} {\cal S}(\alpha_i)~, \nnb \\ \nnb \\
\lla V(q,\lambda) \vel \bar{q}_1(x) g_s \widetilde{G}_{\mu\nu} (ux) 
\gamma_5 q_2(0) \ver 0 \rra \es
-i f_V^T m_V (\varepsilon_\mu^\lambda q_\nu - \varepsilon_\nu^\lambda q_\mu)
\int {\cal D}\alpha_i e^{i(\alpha_{\bar{q}}
+ u \alpha_g) q\mcdot x} \widetilde{{\cal S}}(\alpha_i)~, \nnb \\ \nnb \\
\lla V(q,\lambda) \vel \bar{q}_1(x) g_s \widetilde{G}_{\mu\nu} (ux) 
\gamma_\alpha \gamma_5 q_2(0) \ver 0 \rra \es
f_V m_V q_\alpha (\varepsilon_\mu^\lambda q_\nu - \varepsilon_\nu^\lambda q_\mu)
\int {\cal D}\alpha_i e^{i(\alpha_{\bar{q}}
+ u \alpha_g) q\mcdot x} {\cal A}(\alpha_i)~, \nnb \\ \nnb \\
\lla V(q,\lambda) \vel \bar{q}_1(x) g_s G_{\mu\nu} (ux) i \gamma_\alpha 
q_2(0) \ver 0 \rra \es
f_V m_V q_\alpha (\varepsilon_\mu^\lambda q_\nu - \varepsilon_\nu^\lambda q_\mu)
\int {\cal D}\alpha_i e^{i(\alpha_{\bar{q}}
+ u \alpha_g) q\mcdot x} {\cal V}(\alpha_i)~,
\eaeeq
where $\widetilde{G}_{\mu\nu} = (1/2) \epsilon_{\mu\nu\alpha\beta}
G^{\alpha\beta}$ is the dual gluon field strength tensor, and $\int {\cal D}
\alpha_i = \int d\alpha_q d\alpha_{\bar{q}} d\alpha_g \delta (1 - \alpha_q -
\alpha_{\bar{q}} - \alpha_g)$. 

\eAPP

\newpage

\section*{Figure captions}
{\bf Fig. (1)} The dependence of the strong coupling constant $g_1$
of $\rho^0$ meson to $\Delta^{ +}$  baryon
on Borel mass $M^2$ for several fixed values of the continuum threshold $s_0$. \\ \\
{\bf Fig. (2)} The same as Fig. (1) but for $g_2$. \\ \\
{\bf Fig. (3)} The same as Fig. (1) but for $g_3$. \\ \\
{\bf Fig. (4)} The same as Fig. (1) but for $g_4$. \\ \\
\newpage
\begin{figure}
\vskip 3. cm
    \includegraphics{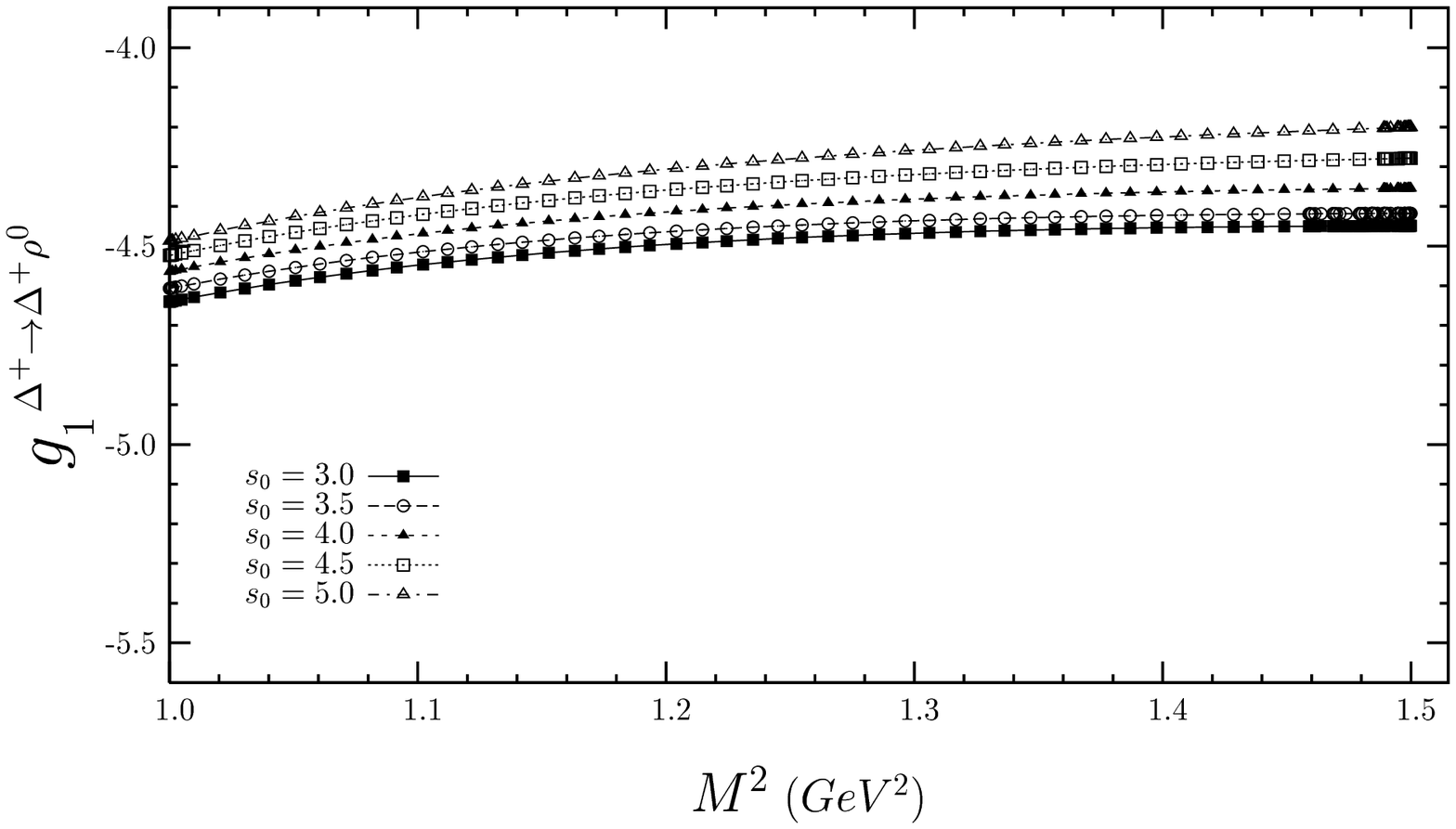}
\vskip 6.3cm
\caption{}
\end{figure}

\begin{figure}
\vskip 4.0 cm
    \includegraphics{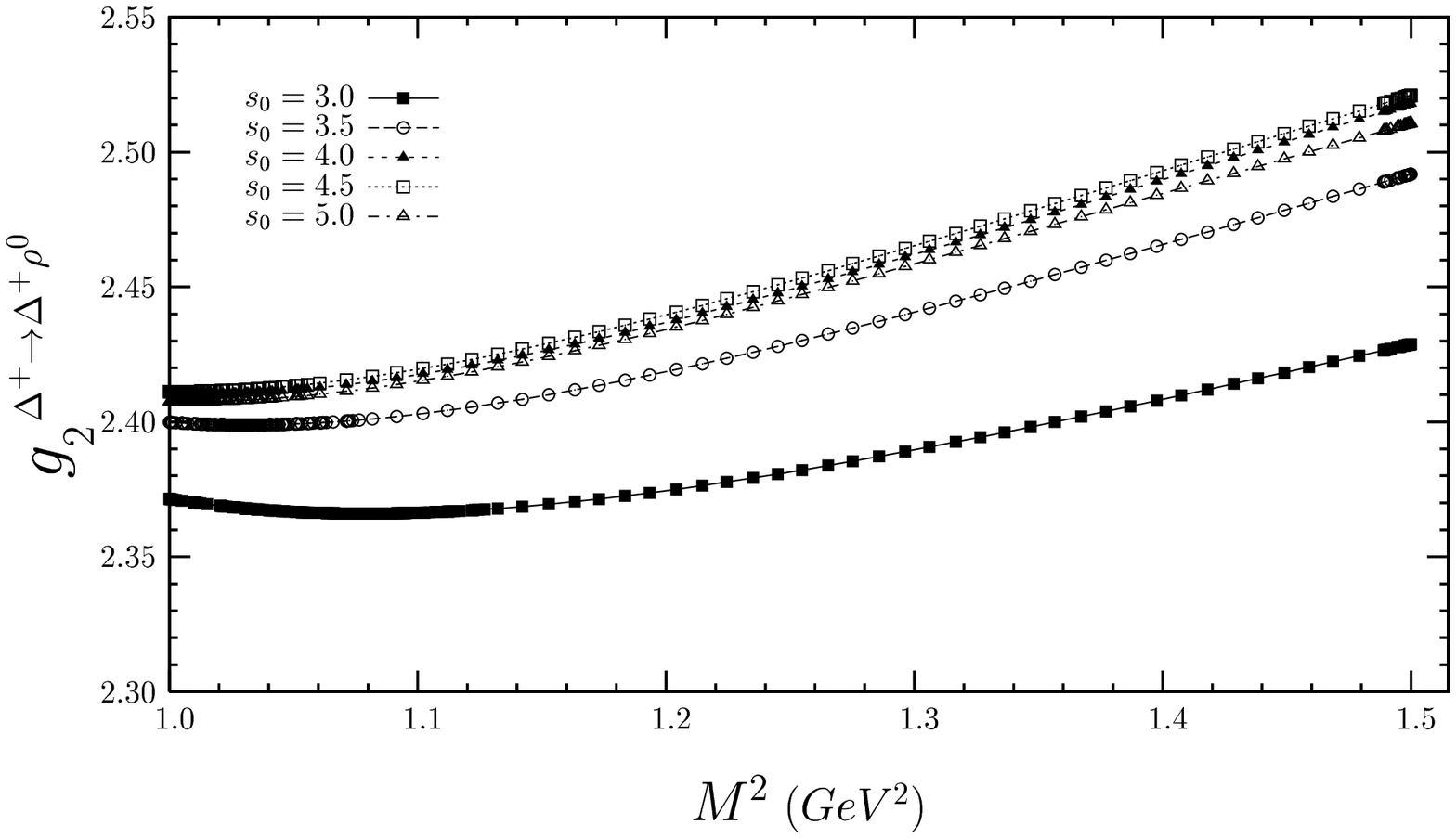}
\vskip 6.3 cm
\caption{}
\end{figure}
\begin{figure}
\vskip 3. cm
    \includegraphics{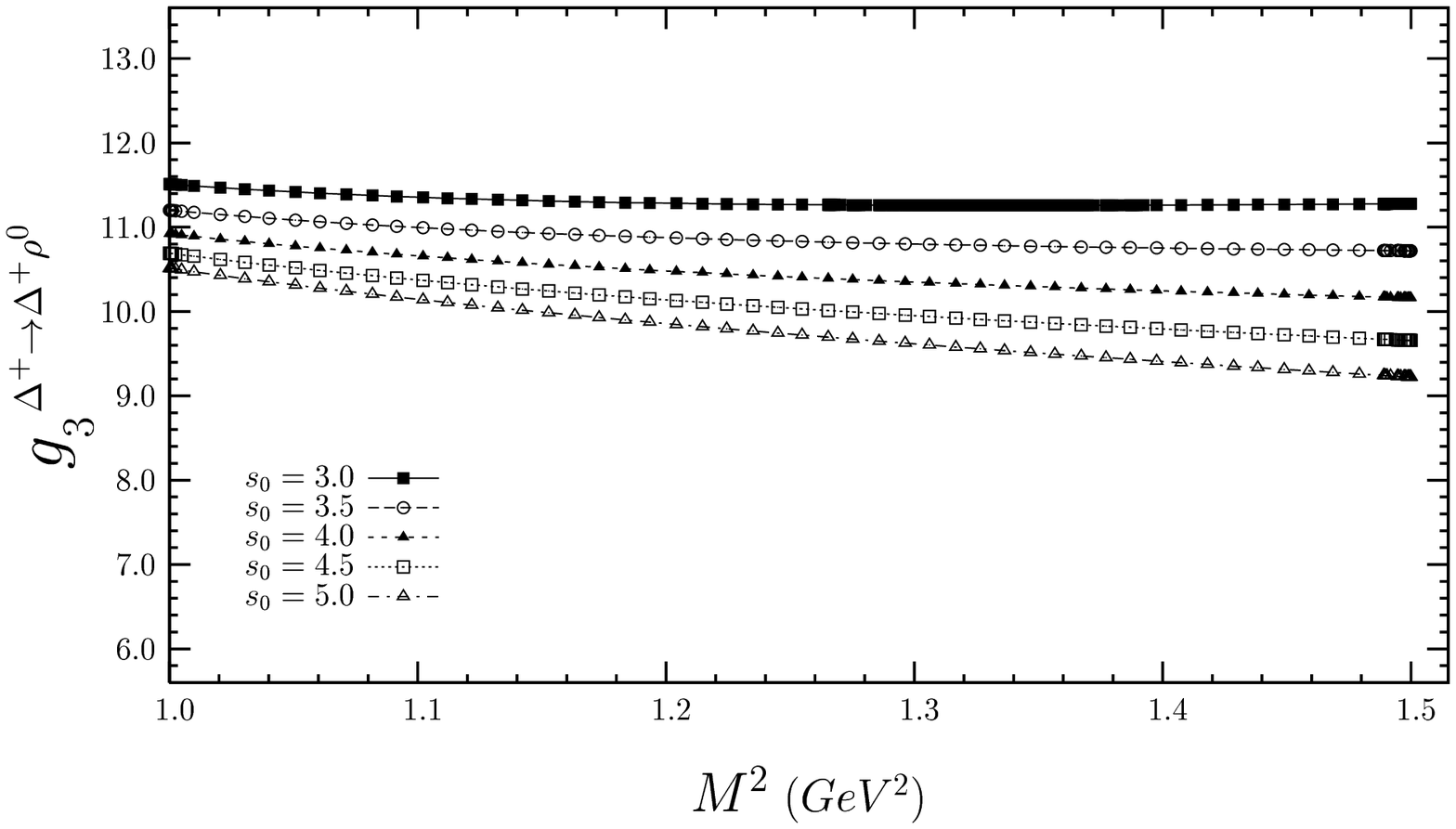}
\vskip 6.3cm
\caption{}
\end{figure}

\begin{figure}
\vskip 4.0 cm
    \includegraphics{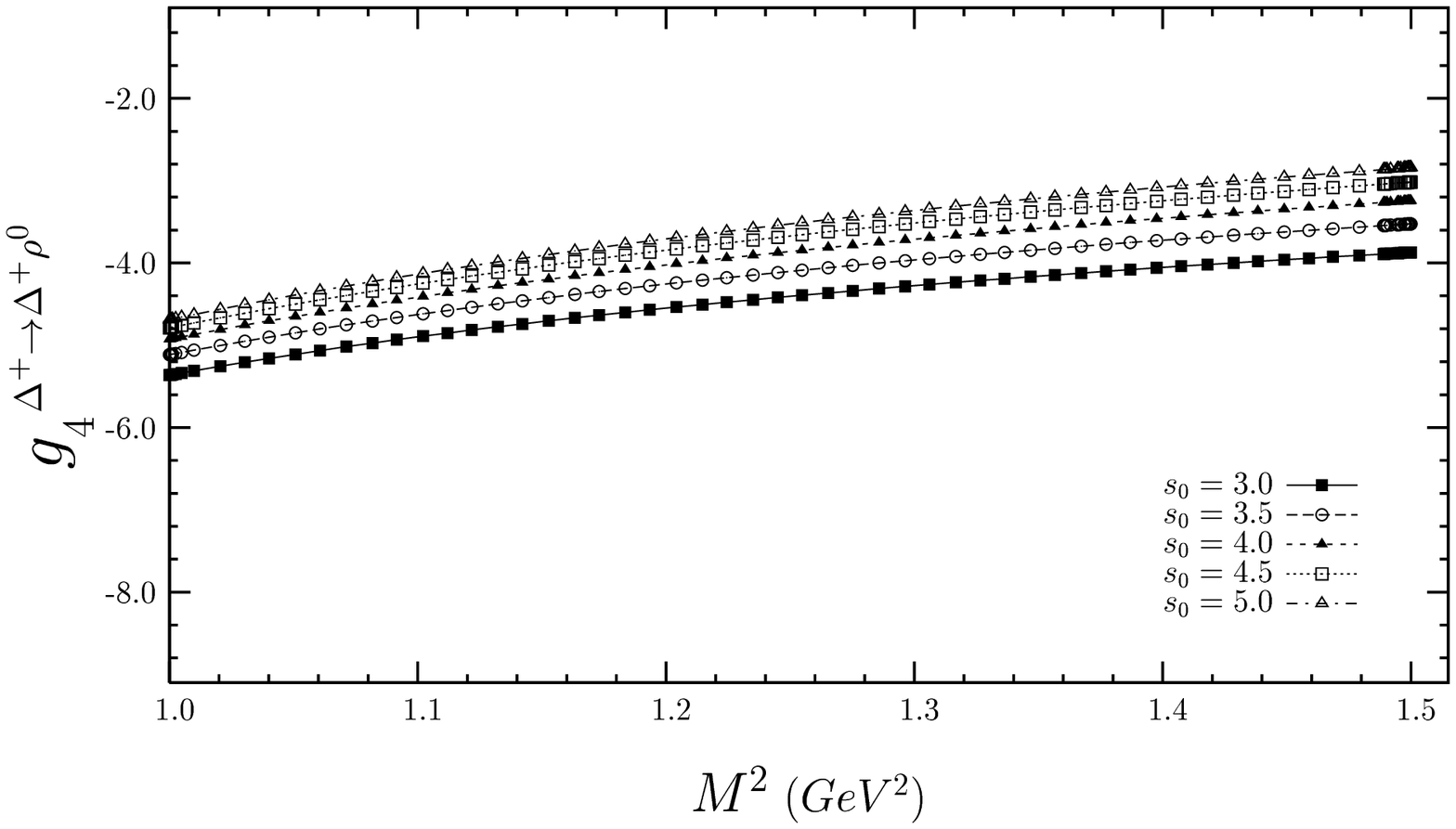}
\vskip 6.3 cm
\caption{}
\end{figure}


\newpage


\begin{thebibliography}{99}

\bibitem{R10501} M. A. Shifman, A. I. Vainshtein and V. I. Zakharov,
  Nucl. Phys. B {\bf 147}, 385 (1979).
\bibitem{R10502} V. M. Braun,
  arXiv: hep--ph/9801222 (1998).



\bibitem{R10503} T. M. Aliev, A. \"{O}zpineci, S. B. Yakovlev, V. Zamiralov,
  Phys. Rev. D {\bf 74}, 116001 (2006).

\bibitem{R10504} T. M. Aliev, A. \"{O}zpineci, M. Savc{\i} and V. Zamiralov,
  Phys. Rev. D {\bf 80}, 016010 (2008).

\bibitem{R10505} T. M. Aliev, K. Azizi, A. \"{O}zpineci and M. Savc{\i},
  Phys. Rev. D {\bf 80}, 096003 (2009).

\bibitem{R10506} T. M. Aliev, A. \"{O}zpineci, M. Savc{\i} and V. Zamiralov,
  Phys. Rev. D {\bf 81}, 056004 (2010).
\bibitem{R10506kazem} T. M. Aliev, K. Azizi, M. Savc{\i}, 	arXiv:1003.5467  [hep-ph].






\bibitem{R10110} P. Ball, V. M. Braun, Y. Koike and K. Tanaka, 
  Nucl. Phys. B {\bf 529}, 323 (1998).

\bibitem{R10111}
  P. Ball, V. M. Braun,
  Nucl. Phys. B {\bf 543}, 201 (1999);
  P. Ball, V. M. Braun,
  Phys. Rev. D {\bf 54}, 2182 (1996).

\bibitem{R10112}
  P. Ball, V. M. Braun, and A. Lenz,
  JHEP 0708.90 (2007).

\bibitem{R10113} I. I. Balitsky, V. M. Braun, and A. V. Kolesnichenko,
  Nucl. Phys. B {\bf 312}, 509 (1989);
  K. G. Chetyrkin, A. Khodjamirian, and A. A. Pivovarov,
  Phys. Lett. B {\bf 651}, 250 (2008).

\bibitem{R10114} T. M. Aliev, A. \"{O}zpineci, M. Savc{\i},
  Phys. Rev. D {\bf 64}, 291 (2006).
\bibitem{R10508} V. M. Belyaev and B. L. Ioffe,
  Sov. Phys. JETP, {\bf 57}, 716 (1982).
\bibitem{R10518} F. X. Lee,
  Phys. Rev. C {\bf 57}  (1998) 322.

\bibitem{R10519} T. M. Aliev, A. \"{O}zpineci, M. Savc{\i},
  Phys. Rev. D {\bf 64} (2001)  034001.
\bibitem{Ioffeb} B. L. Ioffe, Prog. Part. Nucl. Phys. {\bf56} (2006) 232.

\bibitem{Dominguez} C. Dominguez, N. F. Nasrallah, R. Rontisch, K. Schilcher, JHEP {\bf0805}  (2008) 020.




\end{thebibliography}
\end{document}